\documentclass[10pt]{article}  

\usepackage[table]{xcolor}
 \usepackage{amssymb}
\usepackage{mathtools}
\usepackage{amsmath}
\usepackage{verbatim}
\usepackage{soul}
\usepackage{ifthen}
\usepackage{xkeyval}
\usepackage{xcolor}
\usepackage{tikz}
\usepackage{calc}
\usepackage{graphicx}
\usepackage{fancyheadings}
\usepackage{fancyhdr}
\usepackage{booktabs}
\usepackage{multirow}
\usepackage{cite}
\usepackage{enumerate}
\usepackage{authblk}

\usepackage{epstopdf,epsfig}
\usepackage[margin=1in]{geometry}
\DeclareMathOperator{\trace}{trace}
\DeclareMathOperator{\vex}{vex}

\DeclareMathOperator{\SO(3)}{SO(3)}

\begin{document}

\title{Nonlinear Attitude Filtering: A Comparison Study}

\author[1]{M. Zamani,\footnote{Research Associate, Faculty of Electrical Engineering, University of Porto, Email: m.zammani@gmail.com.}}
\affil[1]{University of Porto, Porto, Portugal}
\author[2]{J. Trumpf\footnote{Fellow, Research School of Engineering, The Australian National University. Email: jochen.trumpf@anu.edu.au.}} 
\author[2]{R. Mahony\footnote{Professor, Research School of Engineering, The Australian National University. Email: robert.mahony@anu.edu.au.}}
\affil[2]{The Australian National University, Canberra, ACT, Australia}

\maketitle
\begin{abstract}
This paper contains a concise comparison of a number of nonlinear attitude filtering methods that have attracted attention in the robotics and aviation literature.
With the help of previously published surveys and comparison studies,  the vast literature on the subject is narrowed down to a small pool of competitive attitude filters.
Amongst these filters is a second-order optimal minimum-energy filter recently proposed by the authors.
Easily comparable discretized unit quaternion implementations of the selected filters are provided.
We conduct a simulation study and compare the transient behaviour and asymptotic convergence of these filters in two scenarios with different initialization and measurement errors inspired by applications in unmanned aerial robotics and space flight.
The second-order optimal minimum-energy filter is shown to have the best performance of all filters, including the industry standard multiplicative extended Kalman filter (MEKF).
\end{abstract}

\section{INTRODUCTION}

There are many highly-regarded attitude estimation methods that have been proposed in the literature~\cite{Crassidis,MarkleyBook}. 
In this paper we will focus on recursive, continuous-time filters and exclude non-recursive attitude determination algorithms (e.g.~\cite{Wahba,Shuster1981,Bar1985,Shuster1990,SanyalAttitude2006,Sanyal2008}) , H$_\infty$ filters (e.g.~\cite{MarkleyHinf,MarkleyHinf94,Sanyal2008}), discrete-update filters (e.g.~\cite{2013_Barrau_cdc}), and attitude filters that are based on dynamics or second-order kinematics  (e.g.~\cite{SanyalAttitude2006,Sanyal2008,IzadiSanyal14}).
Within the class of algorithms that we do consider is the industry standard Multiplicative Extended Kalman Filter (MEKF~\cite{markley2003attitude}).
Recent competitors to the MEKF are the Right-Invariant Extended Kalman Filter (RIEKF) that is also known as the Generalized Multiplicative Extended Kalman Filter GMEKF~\cite{GMEKF}, the UnScented QUaternion Estimator  (USQUE~\cite{crassidis2003unscented}) as well as the Geometric Approximate Minimum-Energy (GAME)~\cite{Zamani_TAC12} filter that is a second-order optimal minimum-energy attitude filter recently published by the authors.
In addition, to these variable-gain filters we include the nonlinear complementary attitude observer~\cite{MahonyTac}, that we refer to as the Constant Gain Observer (CGO).
Due to the natural academic process of incremental development of algorithms, it can be difficult to determine what is the state-of-the-art version of any given algorithm.
Moreover, different algorithms use different notation, and even different attitude representations, making cross comparison of competing algorithms difficult.
As a result there is a lack of comparative studies in the literature that provide relative advantages and disadvantages of these methods compared against each other.
Many of the advanced attitude filtering methods are still being advertised by demonstrating performance gain versus a naive implementation of the extended Kalman filter (EKF~\cite{Jazwinski,Anderson}),  an outdated attitude filter with well known convergence issues (cf.~\cite{Choukroun}).

In this paper we document a comprehensive simulation study that sets out to compare the performance of state-of-the-art recursive attitude filters based on the continuous-time attitude kinematics model. The continuous-time selection criterion enables us to capture a larger pool of attitude filters for comparison study as opposed to discrete-time or continuous-discrete-time (discrete-update) attitude filtering that is employed in only a few filters, eg.~\cite{2013_Barrau_cdc,markley2003attitude}.  
We are motivated by the goal of demonstrating the relative performance of the second-order optimal minimum-energy attitude filter recently published by the authors \cite{Zamani_TAC12}, however, we have taken this opportunity to provide a consistent exposition of the structure, merits and performance of the different filters common in the literature.
In particular, the paper presents
\begin{itemize}
\item A literature survey that highlights the state-of-the-art recursive, continuous-time attitude filtering methods that are used in robotics applications.
\item Simple discretized unit quaternion implementations of the state-of-the-art attitude filters in consistent language.
    MATLAB code for the simulation study is available from the fist author's web page:  https://sites.google.com/site/mohammadzamanishomepage.
\item A comprehensive Monte-Carlo simulation study comparing the state-of-the-art filters.  Two scenarios are considered; the first derived from typical parameters associated with attitude estimation for applications in unmanned aerial vehicles (UAVs), and the second from applications in the satellite attitude estimation problem.
 \end{itemize}
We also provide a discussion of how the different algorithms are tuned to obtain best performance.
A key aspect of the simulation studies is the inclusion of gyroscope bias estimation in the estimator algorithms.
In practice, high performance attitude estimation requires on-line bias estimation of the gyroscope, and we have also found that it is a critical factor when comparing relative performance of attitude filters. In fact, in the absence of the bias estimator  most of the state-of-the-art algorithms have comparable performance. We will demonstrate this point using a simulation instance where no bias estimator is present in the competing filters. 

The remainder of the paper is organised as follows.
Section~\ref{Sec_Game} briefly explains attitude filtering and recapitulates the authors' recent GAME filter, a second-order optimal minimum-energy attitude filter.
In Section~\ref{Sec_table} a summary of some of the important attitude filtering methods is provided that also explains the choice of methods that are included in the simulation study.
Section~\ref{Sec_Algorithms} describes the numerical implementation of filters considered in the simulation study.
In particular, the discretization details of each method are provided separately.
In Section~\ref{Sec_Sim} two comparison studies are considered, a UAV simulation setup and a spacecraft simulation setup for which the performance of the GAME filter is compared against the other attitude filters.
Section~\ref{Sec_conclusion} provides the conclusions of the paper.

\section{Attitude Filtering Methods}\label{Sec_table}

This section includes a brief review of some the most important attitude filtering methods that are employed in aerial robotics. In particular, the main ideas behind some attitude filters are explained and a number of these methods are selected, against which the performance of the GAME filter is studied in simulations.

\begin{table}[!]
\resizebox{0.9\textwidth}{!}{\begin{minipage}{1.2\textwidth}
\begin{tabular}{|m{1.35cm}|m{2.35cm}|m{.8cm}|m{3.7cm}|m{10cm}|}
\hline
\hline
\cellcolor[gray]{0.6}Attitude Filters &\cellcolor[gray]{0.6}\hspace{.6cm} state &\cellcolor[gray]{0.6} \hspace{1mm}Ref. & \cellcolor[gray]{0.6}Compared Against  &\cellcolor[gray]{0.6} \hspace{2cm}Comments \\ \hline
\cellcolor{blue!7}GAME & $\SO(3)$ (Unit Quaternions in Simulations) & \hspace{-1mm}Table ~\ref{Sim_Table_GAME} & MEKF, USQUE, Constant Gain Observer & The GAME filter is a 2nd-order approximation to
a minimum-energy filter derived directly on $\SO(3)$ that estimates the gyro bias quickly and is more robust to different noise levels with minimal tuning. \\ \hline
\cellcolor{blue!7} MEKF & Unit\hspace{1cm} Quaternions &~\cite{markley2003attitude} & USQUE, SR-QCKF ~\cite{cubature} & The MEKF estimates a unit quaternion by implicitly
 running an EKF in the vector space of its angular
 velocity input.
 \\ \hline
\cellcolor{blue!7}RIEKF GMEKF & Unit\hspace{1cm} Quaternions &~\cite{GMEKF} & MEKF & RIEKF is a right-invariant construction of the EKF, by considering measurement noise modeled in the inertial frame. The RIEKF has better convergence properties than the MEKF.
\\ \hline
 \cellcolor{blue!7}USQUE & Unit\hspace{1cm} Quaternions & ~\cite{crassidis2003unscented} & MEKF, SR-QCKF ~\cite{cubature}, EKF~\cite{crassidis2003unscented}, BAF~\cite{cheng2004particle} & A three-component attitude error is used to derive an
 unscented filter and the resulting estimated error is
 converted back to unit quaternions and multiplied
 with the previously estimated quaternion to
 produce the filter's estimate.
 \\ \hline
\cellcolor{blue!7}BAF & Unit\hspace{1cm} Quaternions &~\cite{cheng2004particle} & USQUE ~\cite{cheng2004particle} & The BAF achieves comparable performance
 to the USQUE, with the computational costs
of particle filtering
\\ \hline
\cellcolor{blue!7}SR-QCKF & Normalized
 Quaternions & ~\cite{cubature} & USQUE, MEKF ~\cite{cubature} & The USQUE requires more computation
than the SR-QCKF but outperforms it
in mean square error.
\\ \hline
\cellcolor{blue!7}AEKF & Normalized
  Quaternions & ~\cite{markley2004multiplicative} & MEKF & AEKF is conceptually simpler than the MEKF,
but with higher computational cost. The MEKF
is also preferred as it avoids the embedding errors.
\\ \hline
\cellcolor{blue!7}CGO & Unit\hspace{1cm} Quaternions & ~\cite{MahonyTac} &  & A carefully tuned constant gain is used with the same
observer equations as in the MEKF or the GAME filter. It is very robust and asymptotically convergent with minimal computational load but requires exact tuning.
\\ \hline
\cellcolor{blue!7}EKF & Normalized
Quaternions & ~\cite{Anderson} & USQUE ~\cite{crassidis2003unscented} & The EKF in its standard form is outperformed
by the USQUE. The AEKF and the MEKF build up on the EKF to improve the
performance.
\\ \hline
\end{tabular}
\caption{GAME filter: Geometric Approximate Minimum Energy filter, a second-order optimal minimum-energy filter,  MEKF: Multiplicative Extended Kalman Filter, RIEKF: Right-Invariant Extended Kalman Filter, GMEKF: Generalized Multiplicative Extended Kalman Filter,  SR-QCKF: Square-Root Quaternion Kalman
Filter, USQUE: Unscented Quaternion Kalman Filter, BAF: Bootstrap Attitude Filter, AEKF: Additive Extended Kalman Filter, CGO: Constant Gain Observer, EKF:Extended Kalman Filter.}
\label{Sim_Table1}
\end{minipage}}
\end{table}

There are too many attitude filtering methods documented in the literature that we could hope to provide a detailed survey of all such methods in this paper.
A good survey of early attitude filters is given in~\cite{Lefferts82} while more recent material is provided in~\cite{Crassidis}. The recent book~\cite{MarkleyBook}  adds more current methods and detailed explanations, and we will try to cover more recent work on continuous-time recursive algorithms in the present paper.
However, particular applications may require specific modifications or variations of filtering algorithms that, although important for that specific situation, are not interesting in the context of making a more general comparison.
In this paper we will concentrate on our best understanding of what is the most generic algorithm for each filter architecture.

Crassidis~\emph{et.~al} \cite{Crassidis} concluded that ``Many nonlinear filtering methods have been applied to the problem of spacecraft attitude determination in the past three decades. This paper has provided a survey of the methods that its authors consider to be most promising. It remains the case, however, that the extended Kalman filter, especially in the form known as the multiplicative extended Kalman filter, remains the method of choice for the great majority of applications.''.
It remains the case that the MEKF~\cite{markley2003attitude} is an industry standard in recursive attitude filtering and it is an obvious benchmark for the comparisons undertaken in the present paper.
The idea behind the MEKF is to consider the true attitude state as the product of a reference quaternion and an error quaternion that represents the difference between the reference attitude and the true attitude.
The error quaternion is parameterized by a three dimensional representation of attitude and is estimated using an EKF.
The MEKF estimates the true attitude by multiplying the estimated error quaternion (converted back to a unit quaternion) and the reference quaternion \cite{Crassidis}.
In order to avoid the redundancy of having to estimate both the reference quaternion and the error quaternion, the reference quaternion is chosen in a way that the error quaternion is the identity quaternion.
Therefore, the MEKF directly calculates the reference quaternion as a unit quaternion estimate of the true attitude by implicitly running an EKF in the vector space of its angular velocity input.

 The standard EKF, derived in Euler angle local coordinates \cite{Anderson}, is known to have poor performance and stability issues \cite{Choukroun} and
is outperformed by the MEKF~\cite{markley2003attitude}.  The Additive Extended Kalman Filter (AEKF) \cite{markley2004multiplicative} considers the unit quaternion representation of attitude, but initially ignores the unit norm constraint and has been shown to have no better performance than the MEKF~ \cite{markley2004multiplicative}. There are many other clever implementations of the the EKF addressing the attitude filtering problem. However, in our simulation study we only intend to consider  straightforward implementations of the selected mainstream methodologies.  A noteworthy family of attitude filters use a third-degree spherical-radial cubature integration rule to improve the numerical computation of Gaussian weighted integrals. A recent variant, the Square-Root Quaternion Cubature Kalman Filter (SR-QCKF)~\cite{cubature},  offers improved numerical stability by guaranteeing the positiveness of the covariance matrix.

The MEKF is in fact a special case of a more general filter design paradigm termed the left invariant extended Kalman filter IEKF~\cite{InvEKF,bonnabel2009invariant}.
The invariant extended Kalman filter modifies the EKF equations by using an invariant output error rather than a linear error and also by updating the gain using an invariant state error instead of a linear state error.
The right invariant EKF (RIEKF) or generalized MEKF (GMEKF) \cite{GMEKF} is a closely related observer that uses the other-handed invariance for the filter derivation.
The RIEKF is based on the assumption that the state and the output errors are configured in the inertial frame rather than the body-fixed frame, and although this assumption may in itself be questionable, it leads to better stability and conditioning of the associated Riccati equation.
We include the GMEKF in the simulation study to verify its performance improvement over the equivalent IEKF and MEKF algorithms.

The unscented quaternion estimator  (USQUE~\cite{crassidis2003unscented}) is an attitude filter based on the unscented filter (UF~\cite{Unscented}) that has considerable support in the literature and has been proven to work well in many applications.
The UF uses a carefully chosen set of sigma points to approximate the probability distribution as opposed to the EKF that uses local Gaussian noise distributions.
For a naive implementation of the UF, the updated quaternion mean would be obtained by an averaging process that would not in general maintain the unit norm condition of the unit quaternion representing the attitude.
This is overcome in the USQUE~\cite{crassidis2003unscented} where a three-component attitude error is used to derive an unscented filter and the resulting estimated error is converted back to unit quaternions and multiplied with the previously estimated quaternion to produce the attitude estimate.
The hope is that the singularities in the error representation will never occur since the quaternion error is small.
In a recent paper~\cite{cubature}, it was shown that the USQUE had a similar estimation error compared to the MEKF although with a faster convergence rate.
Due to its relatively strong following in the literature and the results in the recent paper \cite{cubature}, the USQUE has been included in the simulation study.
Closely related to the USQUE filter is the Bootstrap Attitude Filter (BAF). This method is based on particle filtering where different to unscented filtering  the samples or particles are drawn randomly to approximate the entire underlying distribution and not just the first two moments. The BAF has been shown~\cite{cheng2004particle} to achieve comparable results to the USQUE , albeit with the high computational load of particle filtering and hence is not selected for our simulation study. 

The second-order optimal minimum-energy (GAME) filter comprises the same observer equations as the continuous-time MEKF~\cite{markley2003attitude}, and the other invariant observers \cite{InvEKF,bonnabel2009invariant}.
However, the Riccati equation of the GAME filter includes curvature correction terms and a geometric second order derivative of the output function that are not present in the algorithms based on stochastic principles. These terms come from computing the full second-order information in the propagation equation for the taylor's expansion of the value-function associated with a deterministic optimal filtering problem, rather than relying on the invariance to propagate local covariance estimates.

There is a large class of constant-gain nonlinear observers designed for attitude estimation (cf.~\cite{Salcudean,Thienel,vasconcelos,MahonyTac,Rouchon,bonnabelTAC2009,Lageman,Trumpf2012analysis}) that are also attractive methods to consider, as they are proven to produce asymptotically convergent estimates.
A simple modification of the observer proposed in~\cite{MahonyTac} allows the inclusion of a (constant) matrix gain rather than the scalar gains in the earlier papers.
For applications where robustness and simplicity of an algorithm is critical, constant gain observers are of significant interest and it is natural to include the constant gain observer in the present simulation study.

\section{Attitude filter formulation}\label{Sec_Game}

Recently, the authors proposed the GAME filter~\cite{Zamani_TAC12}, a second-order optimal minimum-energy filter for the kinematics of the attitude of a rigid body.
Here second-order optimality refers to using a second-order Taylor's expansion approximation of the optimal value-function in the derivation of the filter. This section provides a concise summary of the structure of the GAME filter \cite{Zamani_TAC12} as well as introducing the notation used in the paper.

The attitude kinematics of a rigid-body are given by
\begin{equation}\label{SC_Kin}
\dot{X}(t)=X(t)\Omega_\times(t), \;X(0)=X_0.\\
\end{equation}
Here $X$ is an $\SO(3)$-valued state signal with the unknown initial value $X_0$ and $\Omega \in\mathbb{R}^3$ represents the angular velocity of the
moving body expressed in the body-fixed frame.
The lower index operator $(.)_\times:\mathbb{R}^3\longrightarrow \mathfrak{so}(3)$ denotes the skew-symmetric matrix
\begin{equation}
\Omega_{\times}=\left [\begin{array}{ccc}
0 & -\Omega_3 & \Omega_2\\
\Omega_3 & 0 & -\Omega_1\\
-\Omega_2 & \Omega_1 & 0
\end{array}\right].
\label{eq:times}
\end{equation}

A rate-gyro sensor measures the angular velocity;
\begin{equation}\label{SC_angular velocity}
u(t)= \Omega(t)+B_{\Omega}v_{\Omega}(t)+b(t).
\end{equation}
The signals $u \in\mathbb{R}^3$ and $v_{\Omega} \in\mathbb{R}^3$ denote the body-fixed frame measured angular velocity and the input measurement error, respectively.
The coefficient matrix $B_{\Omega}  \in\mathbb{R}^{3\times3}$ allows for different weightings for the components of the unknown input measurement error $v$.
We assume that $B_{\Omega}$ is full rank and hence that $Q_{\Omega} \coloneqq B_{\Omega} B_{\Omega} ^{\top}$ is positive definite.
In the case of the stochastic filters we will think, in a non-rigorous way, of the noise $v_{\Omega}$ as a unit variance Gaussian process where the matrix $Q_{\Omega}$ can be thought of as the covariance of the actual noise process.
A rigourous development would introduce noise processes in the continuous-time model~(\ref{SC_angular velocity}), however, to simplify the development we will simply use a discrete noise process in the discretization model as is standard in the derivation of the stochastic invariant filters \cite{Crassidis,InvEKF,bonnabel2009invariant,GMEKF}. 
For the minimum-energy deterministic filter, the signal $v_{\Omega}$ does not have a stochastic interpretation, it is simply an auxiliary signal in the cost functional.
For the constant gain observer design the noise signal is ignored. 

The signal $b(t)\in\mathbb{R}^3$ is an unknown slowly time-varying bias signal generated from
\begin{equation}\label{SC_bias}
 \dot{b}(t)= B_bv_b(t),\quad b(0)=b_0,
\end{equation}
where $B_b\in\mathbb{R}^{3\times 3}$ is a full rank weighting matrix known from the model with $Q_b\coloneqq B_bB_b^{\top}$ positive definite.  
The signal $v_b\in\mathbb{R}^3$ is a small unknown perturbation that is once again modeled as a stochastic process in the discretization of the system for the stochastic filters, an auxiliary signal for the minimum-energy filter and ignored for the observer design.
The term $b_0\in \mathbb{R}^3$ is an unknown initial bias.

Consider a collection of known direction vectors $\{\mathring{y}_i \} \in\mathbb{R}^3$ in the reference frame.
Measuring these vectors in the body-fixed frame provides partial information about the attitude $X$.
Typically, magnetometers, visual sensors, sun sensors and star trackers are deployed for this purpose.
The following measurement model is used
\begin{equation}\label{SC_ymeasurements}
 y_i(t)= X(t)^{\top}\mathring{y}_i+D_iw_i(t),\;i=1,\cdots,n
\end{equation}
The measurements $y_i\in\mathbb{R}^3$ are measurements of the $\mathring{y}_i$ in the body-fixed frame and the signals $w_i\in\mathbb{R}^3$  are the unknown output measurement errors.  
The coefficient matrix $D_i\in\mathbb{R}^{3\times3}$ allows for different weightings of the components of the output measurement error $w_i$.
Again, assume that $D_i$ is full rank and $R_i\coloneqq D_iD_i^{\top}$ is positive definite.
In this case the noise $w_i$ is straightforward to interpret as a unit variance Gaussian measurement noise.
In the case of minimum energy filtering it is treated as an auxiliary signal despite its stochastic characteristic while for the observer design it is ignored. 

Consider the cost functional
\begin{equation}\label{SC_cost}
\begin{split}
J&(t;\;X_0,\;b_0,\;v_{\Omega}\vert_{[0,\;t]},\;v_b\vert_{[0,\;t]},\;\{w_i\vert_{[0,\;t]}\})=\dfrac{1}{2}\trace\left[(I-X_0)K^{-1}_{X_0}(I-X_0)^{\top}\right] \\
&\makebox[4cm]{} +\dfrac{1}{2}b_0^{\top}K^{-1}_{b_0}b_0+\dfrac{1}{2}\displaystyle\int^{\top}_0 \left(v_{\Omega}^{\top}v_{\Omega}+v_b^{\top}v_b
+\displaystyle\sum_iw_i^{\top}w_i\right)d\tau,
\end{split}
\end{equation}
in which $K_{X_0},K_{b_0}\in\mathbb{R}^{3\times3}$ are symmetric positive definite matrices.
In the case of minimum-energy filtering, the cost~(\ref{SC_cost}) can be thought of as a measure of the aggregate energy stored in the unknown initialisation and measurement signals
of~(\ref{SC_Kin}),~(\ref{SC_angular velocity}),~(\ref{SC_bias}) and~(\ref{SC_ymeasurements}).
In the case of the stochastic filters this cost functional would be related to an expected value of a log-likelihood cost functional, although, the derivations of the stochastic attitude filters considered \cite{Crassidis,InvEKF,bonnabel2009invariant,GMEKF} do not work from first principles.

In practice, the most common representation used for implementation of attitude estimation algorithms is the unit-quaternion representation.
Unit quaternions lead to more efficient and more robust numerical implementations of the algorithms due to the simple renormalization operation to preserve the representation constraint. 
Moreover, unit quaternions do not have the singularity issue that is associated with many other rotation representations. 
Unit quaternions suffer from non-uniqueness of the representation, however, this issue has been well discussed in the literature and does not pose practical issues for a careful implementation~\cite{MarkleyBook}.
Finally, algorithms such as the MEKF or the USQUE require the unit quaternion representation. 
For these reasons, we will use the unit quaternion representation to express and compare all the algorithms considered. 
Details on the unit quaternions, using the notation in this paper, can be found in the appendix of Mahony \textit{et.~al} \cite{MahonyTac}.

The attitude kinematics $\dot{X}=X\Omega_{\times}$,  in the unit quaternions form is
\begin{equation}\label{App_QuatKin}
\dot{q}=\dfrac{1}{2}A(\Omega)q,
\end{equation}
where $\Omega\in\mathbb{R}^3$ and  
 \begin{equation}
 A(\gamma)\coloneqq \left[\begin{array}{cc}0&-\Omega^{\top}\\ \Omega&-\Omega_{\times}\end{array}\right].
 \end{equation}
The vectorial measurements $y_i =X^{\top}\mathring{y}_i+D_iw_i $, can 
be written as
\begin{equation}\label{App_yq}
y_i = q^{-1}\otimes\bold{p}(\mathring{y}_i)\otimes q+D_iw_i,
\end{equation}
in quaternion notation.  
Here $\bold{p}(\Omega) = (0, \Omega)^\top$ with inverse projection $\bold{p}^{\dagger}$ such that $\bold{p}^{\dagger}(\bold{p}(\Omega)) = \Omega$.

The basic structure of the continuous-time estimation algorithms considered is
\begin{subequations}\label{App_GAME}
 \begin{equation}\label{App_optobs}
\dot{\hat{q}}=\dfrac{1}{2}A\left(u-\hat{b}-P_a\Delta\right)\hat{q},
 \end{equation}
 where  $\hat{b}$ is the estimate of the bias $b$ given from
\begin{equation}\label{App_optbobs}
 \dot{\hat{b}}=P_c^{\top}\Delta,
\end{equation}
with initial conditions $\hat{b}(0)=0$, $\hat{q}(0)=\bold{1}$. 
The innovation term $\Delta$ is defined as
\begin{equation}
\Delta \coloneqq\displaystyle\sum_i(R^{-1}_i(\hat{y}_i-y_i))\times \hat{y}_i.
\end{equation}
where we recall that $R_i\coloneqq D_iD_i^{\top}$. 
The gains $P_a$ and $P_c$ are symmetric $3\times3$ matrices updated from a Riccati like equation. 
For example, in the case of the GAME filter \cite{Zamani_TAC12}, and recalling $Q_{\Omega}\coloneqq BB^{\top}$, $Q_b\coloneqq B_bB_b^{\top}$, the Riccati equation in continuous time is written \cite{Zamani_TAC12}
\begin{equation}\label{App_briccati}
\begin{split}
&\dot{P}_a=Q_{\Omega}+2\mathbb{P}_s(P_a(2(u-\hat{b})-P_a\Delta)_{\times})+P_a(E-S)P_a-P^{\top}_c-P_c,\\
&\dot{P}_c = -(u-\hat{b}-P_a\Delta)_{\times}P_c+P_a(E-S)P_c-P_b,\\
\end{split}
\end{equation}
where the gain $P_b$ is given from
\begin{equation}\label{App_P_b}
\dot{P}_b = Q_b +P_c(E-S)P_c, 
\end{equation}
\end{subequations}
with 
\begin{equation}\label{App_ES}
S\coloneqq \sum_i(\hat{y}_i)^{\top}_{\times}R^{-1}_i(\hat{y}_i)_{\times},\;E\coloneqq\trace(C)I-C,\;C\coloneqq\sum_i\mathbb{P}_s(R^{-1}_i(\hat{y}-y_i)\hat{y}^{\top}_i).
\end{equation}
Here $P_a(0)=(\trace(K_{X_0})I-K_{X_0})^{-1}$ where $K_{X_0}$ is a known variable from the cost function.
The vector $\hat{y} = \hat{q}^{-1}\otimes\bold{p}(\mathring{y}_i)\otimes\hat{q}$ is analogous to to~(\ref{App_yq}) and the operator $\mathbb{P}_s$ is a symmetric projector defined in equation~(\ref{Not_SymProj}) below.

The matrix 
\[
P = \begin{pmatrix} P_a & P_b \\ P_b^\top & P_c \end{pmatrix}
\]
contains the full gain matrix information including both the attitude and bias states. 
It is an interesting feature of all the attitude filtering algorithms that the matrix $P_b$ is not required to implement the filter update equations 
(\ref{App_optobs}) and (\ref{App_optbobs}). 
It is of course, coupled in the differential Riccati equation (\ref{App_briccati}) and (\ref{App_P_b}) and must be computed. 
However, it is natural to write the Riccati equation in decoupled form with separate matrices $P_a$, $P_b$, $P_c$. 
Although equations (\ref{App_GAME}) are specific to the GAME filter, the same notation and structure has been used for all the filters in this paper. 

In order to numerically compare the filters, discrete-time implementations of the continuous-time filters are required.  
This is not a trivial task in attitude filtering (and similar problems) as the Lie group configuration of the underlying state space has to be preserved during the numerical computation.
In practice, at least when the time step considered is small compared to the motion of the vehicle, the usual approach taken is to use a simple Lie group Euler numerical integration.
Assuming a small time step $dt$, then $\Omega$ is approximately constant on the time period $[kdt,\;(k+1)dt]$ for $k\in\mathbb{N}$.
Denote this value by $\Omega_k$. 
Then the exact integration of~(\ref{App_QuatKin}) yields
\begin{equation}\label{Sim_Quatint}
q_{k+1}=\dfrac{1}{2}\exp(dtA(\Omega_k))q.
\end{equation}
This numerical integrator is used for the attitude state (\ref{App_optobs}). The bias observer~(\ref{App_optbobs}) as well as the Riccati equations~(\ref{App_briccati}) and~(\ref{App_P_b}) live in linear vector spaces and are implemented using a classical Euler iteration. The same simple numerical integration scheme is used for all the filters in this paper since the focus is on a comparison of the performance of the different filter architectures and not on the numerical implementation.

\section{Discretized Implementations of the Attitude Filters Selected for the Simulation Study}\label{Sec_Algorithms}
In this section unit quaternion discretized implementations of the attitude kinematics~(\ref{SC_Kin}) and the filters selected for our simulation study are provided in Tables~\ref{Sim_Table_system} to~\ref{Sim_Table_CGO} below. 
The following notation is instrumental in these formulations.

 Recall that the set of skew symmetric matrices is denoted by $\mathfrak{so}(3)$ and is the Lie-algebra of $\SO(3)$. The operation $(\cdot)_\times$ is given by (\ref{eq:times}) and provides the isomorphism between $\mathbb{R}^3$ and $\mathfrak{so}(3)$.
The inverse operator $\vex:\mathfrak{so}(3)\longrightarrow \mathbb{R}^{3}$ extracts the skew coordinates, $\vex(\Omega_\times)=\Omega$.

Denote a symmetric positive semi-definite matrix by $B\geq 0$ (a symmetric positive definite matrix is denoted by $B>0$).
The seminorm $\Vert . \Vert_{R}: \mathbb{R}^{3\times 3}\longrightarrow\mathbb{R}_0^{+}$ is given by
\begin{equation}\label{wmnorm}
\Vert M \Vert_{R} \coloneqq \sqrt{\dfrac{1}{2}\trace(MR M^{\top})},
\end{equation}
where $R \in\mathbb{R}^{3\times3}\geq0$.
Note that if $R>0$ then $\Vert M \Vert_{R}$ coincides with the Frobenius norm of $MR^{1/2}$.
The symmetric projector $\mathbb{P}_s$ is defined by
\begin{equation}\label{Not_SymProj}
\mathbb{P}_s(M)\coloneqq 1/2(M+M^{\top}).
\end{equation}
The skew-symmetric projector $\mathbb{P}_a$ is defined by
\begin{equation}
\mathbb{P}_a(M)\coloneqq 1/2(M-M^{\top}).
\end{equation}
Note that for every $A\in\mathfrak{so}(3)$, $M\in\mathbb{R}^{3 \times 3}$ and $S=S^{\top}\in\mathbb{R}^{3\times 3}$ ,
\begin{equation}\label{trace}
\trace(A\mathbb{P}_s(M))=0,\;\trace(\mathbb{P}_s(SA))=0.
\end{equation}

\begin{table}[!ph]
\begin{tabular}{|p{4cm}|p{8cm}|}
\hline
\multirow{5}{*}{Kinematics}&\\& $q(k+1)=\frac{1}{2}\exp(dtA(\Omega(k)))q(k)$, $q(0)=q_0$,\\&\\&$A(\Omega(k))=\left[\begin{array}{cc}0&-\Omega(k)^{\top}\\ \Omega(k)&-\Omega(k)_{\times}\end{array}\right]$\\&\\
\hline
\multirow{4}{*}{Gyro Measurements}& \\&$ u(k) = \Omega(k) + b(k) + B_{\Omega}v_{\Omega}(k)$,\\&$Q_{\Omega}= B_{\Omega}B_{\Omega}'$\\&\\
\hline
\multirow{3}{*}{Bias Model}&\\& $b(k+1) = b(k) + dt[B_bv_b(k)],\; Q_b = B_vB_v'$\\&\\
\hline
\multirow{5}{*}{Vector Measurements}&\\& $y_i(k)= \bold{p}^{\dagger}(q(k)^{-1}\otimes\bold{p}(\mathring{y}_i)\otimes q(k))+D_iw_i(k)$,\\
&$R_i =D_iD_i'$,\\&$\bold{p}(\mathring{y}_i) = \left[\begin{array}{cc}0\\\mathring{y}_i\end{array}\right],\;\bold{p}^{\dagger}(\bold{p}(\mathring{y}_i))=\mathring{y}_i$\\&\\
\hline
\end{tabular}
 \caption{Discrete Attitude Kinematics and Measurements}
\label{Sim_Table_system}
\end{table}
\begin{table}[!phtb]
\begin{tabular}{|p{3cm}|p{10cm}|}
\hline
\multirow{9}{*}{Attitude Observer}&\\& $\hat{q}(k+1)=\dfrac{1}{2}\exp(dtA[u(k)-\hat{b}(k)+P_a(k)\Delta(k)])\hat{q}(k)$,\\&\\& $\Delta(k)=\displaystyle\sum_i\hat{y}_i(k)\times(R^{-1}_i(\hat{y}_i(k)-y_i(k))) $,\\&\\&
$\hat{y}_i(k) = \bold{p}^{\dagger}(\hat{q}(k)^{-1}\otimes\bold{p}(\mathring{y}_i)\otimes \hat{q}(k))$,  $\hat{q}(0)=[1\;0\;0\;0]^{\top}$\\&\\
\hline
\multirow{3}{*}{Bias Observer}&\\& $\hat{b}(k+1)=\hat{b}(k)+dt[P_c(k)^{\top}\Delta(k)]$, $\hat{b}(0)=[0\;0\;0]^{\top}$,\\&\\
\hline
\multirow{13}{*}{Riccati Gains}&\\&$P_a(k+1)=P_a(k)+dt[Q_{\Omega}+2\mathbb{P}_s[P_a(k)(u(k)-\hat{b}(k)-\frac{1}{2}P_a(k)\Delta(k))_{\times}-P_c(k)]+P_a(k)(E(k)-S(k))P_a(k)]$,\\&\\&
$P_c(k+1) =P_c(k)+dt[-(u(k)-\hat{b}(k)-\frac{1}{2}P_a(k)\Delta(k))_{\times}P_c(k)+P_a(k)(E(k)-S(k))P_c(k)-P_b(k)]$,\\&\\&
$P_b(k+1) =P_b(k)+dt[Q_b +P_c(k)(E(k)-S(k))P_c(k)],$\\&\\&
$S(k)= \sum_i(\hat{y}_i(k))^{\top}_{\times}R^{-1}_i(\hat{y}_i(k))_{\times},$\\&\\&
$E(k)=\trace(C(k))I-C(k),$\\&\\&$C(k)=\sum_i\mathbb{P}_s(R^{-1}_i(\hat{y}_i(k)-y_i(k))\hat{y}(k)^{\top}_i),$\\&\\
\hline
  \end{tabular}
  \caption{GAME Filter}
\label{Sim_Table_GAME}
\end{table}
 
\begin{table}[!htb]
\begin{tabular}{|p{3cm}|p{10cm}|}
\hline
\multirow{9}{*}{Attitude Observer}&\\& $\hat{q}(k+1)=\dfrac{1}{2}\exp(dtA[u(k)-\hat{b}(k)+P_a(k)\Delta(k)])\hat{q}(k)$,\\&\\& $\Delta(k)=\displaystyle\sum_i\hat{y}_i(k)\times(R^{-1}_i(\hat{y}_i(k)-y_i(k))) $,\\&\\&
$\hat{y}_i(k) = \bold{p}^{\dagger}(\hat{q}(k)^{-1}\otimes\bold{p}(\mathring{y}_i)\otimes \hat{q}(k))$,  $\hat{q}(0)=[1\;0\;0\;0]^{\top}$\\&\\
\hline
\multirow{3}{*}{Bias Observer}&\\& $\hat{b}(k+1)=\hat{b}(k)+dt[P_c(k)^{\top}\Delta(k)]$, $\hat{b}(0)=[0\;0\;0]^{\top}$,\\&\\
\hline
\multirow{11}{*}{Riccati Gains}&\\&$P_a(k+1)=P_a(k)+dt[Q_{\Omega}+2\mathbb{P}_s[P_a(k)(u(k)-\hat{b}(k))_{\times}-P_c(k)]-P_a(k)S(k)P_a(k)]$,\\&\\&
$P_c(k+1) =P_c(k)+dt[-(u(k)-\hat{b}(k))_{\times}P_c(k)-P_a(k)S(k)P_c(k)-P_b(k)]$,\\&\\&
$P_b(k+1) =P_b(k)+dt[Q_b -P_c(k)S(k)P_c(k)],$\\&\\&
$S(k)= \sum_i(\hat{y}_i(k))^{\top}_{\times}R^{-1}_i(\hat{y}_i(k))_{\times},$\\&\\
\hline
  \end{tabular}
  \caption{MEKF}
\label{Sim_Table_MEKF}
\end{table}

\begin{table}[!htb]
 \begin{tabular}{|p{3cm}|p{10cm}|}
\hline
\multirow{9}{*}{Attitude Observer}&\\& $\hat{q}(k+1)=\dfrac{1}{2}\exp(dtA[u(k)-\hat{b}(k)+P_a(k)\Delta(k)])\hat{q}(k)$,\\&\\& $\Delta(k)=\displaystyle\sum_i\mathring{y}_i\times(R^{-1}_i(\mathring{y}_i-\hat{y}_i(k))) $,\\&\\&
$\hat{y}_i(k) = \bold{p}^{\dagger}(\hat{q}(k)\otimes\bold{p}(y_i(k))\otimes \hat{q}^{-1}(k))$,  $\hat{q}(0)=[1\;0\;0\;0]^{\top}$\\&\\
\hline
\multirow{3}{*}{Bias Observer}&\\& $\hat{b}(k+1)=\hat{b}(k)+dt[\hat{q}(k)^{-1}\otimes\bold{p}(P_c(k)^{\top}\Delta(k))\otimes \hat{q}(k)]$, $\hat{b}(0)=[0\;0\;0]^{\top}$,\\&\\
\hline
\multirow{11}{*}{Riccati Gains}&\\&$P_a(k+1)=P_a(k)+dt[Q_{\Omega}-2\mathbb{P}_s[P_c(k)]-P_a(k)S(k)P_a(k)]$,\\&\\&
$P_c(k+1) =P_c(k)+dt[-P_c(k)\hat{q}(k)\otimes\bold{p}((u(k)-\hat{b}(k))_{\times})\otimes \hat{q}^{-1}(k)-P_a(k)S(k)P_c(k)-P_b(k)]$,\\&\\&
$P_b(k+1) =P_b(k)+dt[2\mathbb{P}_s(\hat{q}(k)\otimes\bold{p}((u(k)-\hat{b}(k))_{\times})\otimes \hat{q}^{-1}(k)P_b(k)+Q_b -P_c(k)S(k)P_c(k)],$\\&\\&
$S(k)= \sum_i(\mathring{y}_i(k))^{\top}_{\times}R^{-1}_i(\mathring{y}_i(k))_{\times},$\\&\\
\hline
  \end{tabular}
  \caption{RIEKF (GMEKF)}
\label{Sim_Table_RIEKF}
 \end{table}

\begin{table}[!htb]
  \footnotesize
\resizebox{1.3\textwidth}{!}{\begin{minipage}{1.3\textwidth}
\begin{tabular}{|p{3.2cm}|p{14cm}|}
\hline
\multirow{1}{*}{Parameters}&$\hat{x}^{+}(0)=[\textbf{0}^{\top}\;\hat{b}_0^{\top}]^{\top}$,\;$a=1$,\; $f=4$,\; $\lambda=1$,\; $n=6$,\\
\hline
\multirow{2}{*}{Discrete $Q_k$}&\vspace{.01mm}$Q_k =\frac{dt}{2}\left[\begin{array}{cc}Q_{\Omega}-\frac{dt^2}{6}Q_b &0_{3\times 3},\\0_{3\times 3}&Q_b\end{array}\right]$,\\
\hline
\multirow{1}{*}{Sigma Points}&$\sigma_k\leftarrow 2n\;\mbox{ columns from} \pm\sqrt{(n+\lambda)[P^+_k+Q_k]}$,\quad $\mathcal{X}_k(0)=\hat{x}^+_k$, $\mathcal{X}_k(i)=\sigma_k(i)+\hat{x}^+_k$,\\
\hline
\multirow{2}{*}{Error Quaternions}&$\delta q^+_{4_k}(i)=\frac{-a\Vert\mathcal{X}^{\delta p}_k(i)\Vert^2+f\sqrt{f^2 +(1-a^2)\Vert\mathcal{X}^{\delta p}_k(i)\Vert^2}}{f^2+\Vert\mathcal{X}^{\delta p}_k(i)\Vert^2}$,\quad$\delta\varrho^+_k(i)=f^{-1}[a+\delta q^+_{4_k}(i)]\mathcal{X}^{\delta p}_k(i),$\\
&$\delta q^+_k(i)=[\delta q^+_{4_k}(i)\; \delta\varrho^{+^{\top}}_k(i)]^{\top},\quad i=1,2\cdots,12$,\\
\hline
\multirow{1}{*}{Sigma Quaternions}&$\hat{q}^+_k(0)=\hat{q}^+_k,$,\quad$\hat{q}^+_k(i)=\delta q^+_k(i)\otimes \hat{q}^+_k$,\\
\hline
\multirow{3}{*}{Propagation}&$\hat{q}^-_{k+1}(i)= \dfrac{1}{2}\exp(dtA[u(k)-\mathcal{X}^{\hat{b}}_k(i)])\hat{q}^+_k(i)$,\quad$ i=0,1,\cdots ,12$,\\
&$\delta q^-_{k+1}(i)= \hat{q}^-_{k+1}(i)\otimes (\hat{q}^-_{k+1}(0))^{-1}$,\quad $\delta q^-_{k+1}(0)=[1\;0\;0\;0]^{\top}$,\\
&$[\delta q^-_{4_{k+1}}(i)\; \delta\varrho^{-^{\top}}_{k+1}(i)]^{\top}=\delta q^-_{k+1}(i)$,\quad$\mathcal{X}^{\delta p}_{k+1}(i)=f\frac{\delta\varrho^{-}_{k+1}(i)}{a+\delta q^-_{4_{k+1}}(i)}$, $\mathcal{X}^{\delta p}_{k+1}(0)=\textbf{0}$,\quad$\mathcal{X}^{\hat{b}}_{k+1}(i)=\mathcal{X}^{\hat{b}}_{k}(i),$\\
\hline
\multirow{2}{*}{Prediction}&$\hat{x}^-_{k+1}=\frac{1}{n+\lambda}\left\{\lambda\mathcal{X}_{k+1}(0)+\frac{1}{2}\sum_{i=1}^{2n}\mathcal{X}_{k+1}(i)\right\}$,\\&$P^-_{k+1}=\frac{1}{n+\lambda}\left\{\lambda[\mathcal{X}_{k+1}(0)-\hat{x}^-_{k+1}][\mathcal{X}_{k+1}(0)-\hat{x}^-_{k+1}]^{\top}\right.$ $\left.+\frac{1}{2}\sum_{i=1}^{2n}[\mathcal{X}_{k+1}(i)-\hat{x}^-_{k+1}][\mathcal{X}_{k+1}(i)-\hat{x}^-_{k+1}]^{\top}\right\}+Q_k$,\\
\hline
\multirow{2}{*}{Mean Observations}&$\gamma_{k+1}(i)=\left[\begin{array}{c} \bold{p}^{\dagger}(\hat{q}_{k+1}^-(i)^{-1}\otimes\bold{p}(\mathring{y}_1)\otimes \hat{q}_{k+1}^-(i))\\  \bold{p}^{\dagger}(\hat{q}_{k+1}^-(i)^{-1}\otimes\bold{p}(\mathring{y}_2)\otimes \hat{q}_{k+1}^-(i))\\ \vdots\end{array}\right]$,\quad
$\hat{y}^-_{k+1} = \frac{1}{n+\lambda}\left\{\lambda\gamma_{k+1}(0)+\frac{1}{2}\sum_{i=1}^{2n}\gamma_{k+1}(i)\right\}$,\\
\hline
\multirow{3}{*}{Covariance Update}&$P^{yy}_{k+1}=\frac{1}{n+\lambda}\left\{\lambda[\gamma_{k+1}(0)-\hat{y}^-_{k+1}][\gamma_{k+1}(0)-\hat{y}^-_{k+1}]^{\top}\right.$ $\left.+\frac{1}{2}\sum_{i=1}^{2n}[\gamma_{k+1}(i)-\hat{y}^-_{k+1}][\gamma_{k+1}(i)-\hat{y}^-_{k+1}]^{\top}\right\}$,
\\
&$P^{\nu\nu}_{k+1}=P^{yy}_{k+1}+R_{k+1}$,\\&
$P^{xy}_{k+1}=\frac{1}{n+\lambda}\left\{\lambda[\mathcal{X}_{k+1}(0)-\hat{x}^-_{k+1}][\gamma_{k+1}(0)-\hat{y}^-_{k+1}]^{\top}\right.$ $\left.+\frac{1}{2}\sum_{i=1}^{2n}[\mathcal{X}_{k+1}(i)-\hat{x}^-_{k+1}][\gamma_{k+1}(i)-\hat{y}^-_{k+1}]^{\top}\right\}$\\
\hline
\multirow{1}{*}{Update}&$\hat{x}^+_{k+1}=\hat{x}^-_{k+1} + P^{xy}_{k+1}(P^{\nu\nu}_{k+1})^{-1}(y_{k+1}-\hat{y}_{k+1}),$\quad $P^+_{k+1} = P^-_{k+1} - P^{xy}_{k+1}(P^{\nu\nu}_{k+1})^{-\top}P^{xy}_{k+1}$,\\
\hline
\multirow{2}{*}{Quaternion Update}&$\hat{x}^+_{k+1}= [\delta p^{\top}\;\hat{b}^{+^{\top}}]^{\top}$,\quad
$\delta q^+_{4_{k+1}}=\frac{-a\Vert\delta p\Vert^2+f\sqrt{f^2 +(1-a^2)\Vert\delta p\Vert^2}}{f^2+\Vert\delta p\Vert^2}$,\\
&$\delta\varrho^+_{k+1}=f^{-1}[a+\delta q^+_{4_{k+1}}]\delta p,$\quad$\delta q^+_{k+1}=[\delta q^+_{4_{k+1}}\; \delta\varrho^{+^{\top}}_{k+1}]^{\top}$,\quad$\hat{q}^+_{k+1}=\delta q^+_{k+1}\otimes \hat{q}^-_{k+1}$\\
\hline
\end{tabular}
   \caption{USQUE}
\label{Sim_Table_USQUE}\end{minipage}}
 \end{table}

\begin{table}[!htb]
\begin{tabular}{|p{3cm}|p{10cm}|}
\hline
\multirow{9}{*}{Attitude Observer}&\\& $\hat{q}(k+1)=\dfrac{1}{2}\exp(dtA[u(k)-\hat{b}(k)+k_P\Delta(k)])\hat{q}(k)$,\\&\\& $\Delta(k)=\displaystyle\sum_iy_i(k)\times \hat{y}_i(k)$,\\&\\&
$\hat{y}_i(k) = \bold{p}^{\dagger}(\hat{q}(k)^{-1}\otimes\bold{p}(\mathring{y}_i)\otimes \hat{q}(k))$,  $\hat{q}(0)=[1\;0\;0\;0]^{\top}$\\&\\
\hline
\multirow{3}{*}{Bias Observer}&\\& $\hat{b}(k+1)=\hat{b}(k)-dt[k_I\Delta(k)]$, $\hat{b}(0)=[0\;0\;0]^{\top}$\\&\\
\hline
  \end{tabular}
  \caption{CGO}
\label{Sim_Table_CGO}
 \end{table}
 \clearpage
\section{Comparison Study}\label{Sec_Sim}
In this section, multiple simulated experiments are presented that compare the filtering methods considered in Section~\ref{Sec_table}.\\

\subsection{Case 1: Measurement Errors Expected from  Low-Cost UAV Sensors}
The first experiment considered simulates attitude estimation for a low cost unmanned aerial vehicle (UAV) system for which the measurement errors are relatively large. 
It is also assumed that the rotation and the bias initialization errors are large, as is the case when using low cost MEMS gyros such as the popular InvenSense MPU-3000 family. 
The simulation parameters are summarized in Table~\ref{Sim_Table_UAV}. 
The GAME filter (Table~\ref{Sim_Table_GAME}) is compared against the MEKF (Table~\ref{Sim_Table_MEKF}), the RIEKF (Table~\ref{Sim_Table_RIEKF}), the USQUE (Table~\ref{Sim_Table_USQUE}) and the CGO (Table~\ref{Sim_Table_CGO}) that are explained in detail in Section~\ref{Sec_table}.

Simulated attitude kinematics and measurements (see Table~\ref{Sim_Table_system}) are considered with the following parameters that are also summarized in Table~\ref{Sim_Table_UAV}.
A sinusoidal input $\Omega = [\sin(\frac{2\pi}{15}t)\; -\sin(\frac{2\pi}{18}t+\pi/20)\; \cos(\frac{2\pi}{17}t)]$ drives the true trajectory $q$.
The input measurement errors $v$ and $v_b$ are Gaussian zero mean random processes with unit variance. 
The coefficient matrix $B$ is chosen so that the signal $Bv$ has a standard deviation of $25$ degrees per `second'.
The bias variation is adjusted by $B_b$ such that $B_bv_b $ has a standard deviation of $0.1$ degrees per `second' squared.  
The system is initialized with a unit quaternion representing a rotation with standard deviation of $std_{q_0}=60$ degrees and an initial bias with standard deviation of $std_{b_0}=20$ degrees per `second'. 
We assume that two orthogonal unit reference vectors are available. 
We also consider Gaussian zero mean measurement noise signals $w_i$ with unit standard deviations. 
The coefficient matrices $D_i$ are chosen so that the signals $D_iw_i$ have standard deviations of $30$ degrees.
Although the two filters do not have access to the noise signals $v_{\Omega}$, $v_b$ and $w_i$ themselves, they have access to the matrices $Q_{\Omega}=BB^{\top}$, $Q_b=B_bB^{\top}_b$ and $R_i=D_iD^{\top}_i$.
The filters are simulated using zero initial bias estimates and using the identity unit quaternion as their initial quaternion estimate.

The following filter initializations are considered that are also summarised in Table~\ref{Sim_Table_tun}. 
The initial quaternion and bias gain matrices of the USQUE are chosen according to the variance of the system's initial quaternion in radians $P_a(0)=std^2_{q_0}I_{3\times 3}$ and the variance of the system's initial bias in radians per `seconds' $P_b(0) = std^2_{b_0}I_{3\times 3}$.
The initial quaternion and bias gain matrices of the GAME filter, the MEKF and the RIEKF are chosen according to the inverse variance of the system's initial quaternion in radians $P_a(0)=\frac{1}{std^2_{q_0}}I_{3\times 3}$ and the inverse variance of the system's initial bias in radians per `seconds' $P_b(0) = \frac{1}{std^2_{b_0}}I_{3\times 3}$ as these filters are in the information form. The coupling initial gain is considered as the zero matrix $P_c(0)=0_{3\times 3}$ for all the filters. The CGO is initialized with $k_p  = 1$ and $k_I=0.3$ as in~\cite{MahonyTac}.


\begin{table}[h]
\begin{tabular}{|p{5cm}|p{8cm}|}
\hline
\vspace{.1mm}Time Step &\vspace{.1mm} 0.001 ($s$)\\
\hline
\vspace{.1mm}Simulation Time & \vspace{.1mm}50 ($s$)\\
\hline
\vspace{.1mm}Angle of Rotation Initialization Error  & \vspace{.1mm}$\mathcal{N}\sim(0,60^2)^{\circ}$\\
\hline
\vspace{.1mm}Bias Initialization Error   & \vspace{.1mm}$\mathcal{N}\sim(0,20^2)\frac{^{\circ}}{s}$\\
\hline
\vspace{.1mm}Reference Directions & \vspace{.1mm}$\mathring{y}_1=[1\; 0 \;0]$, $\mathring{y}_2=[0\; 1 \;0]$\\
\hline
\vspace{.1mm}Input signal& \vspace{.1mm}$\Omega = [\sin(\frac{2\pi}{15}t)\; -\sin(\frac{2\pi}{18}t+\pi/20)\; \cos(\frac{2\pi}{17}t)]\frac{rad}{s}$\\
\hline
\vspace{.1mm}Input error $B_{\Omega}v_{\Omega}$ &\vspace{.1mm} $\mathcal{N}\sim(0,25^2)\frac{^{\circ}}{s}$\\
\hline
\vspace{.1mm}Bias Variation $B_bv_b$ &\vspace{.1mm} $\mathcal{N}\sim(0,0.1^2)\frac{^{\circ}}{s^2}$\\
\hline
\vspace{.1mm}Measurement error  $D_iw_i$ &\vspace{.1mm} $\mathcal{N}\sim(0,30^2)^{\circ}$\\
\hline
  \end{tabular}
  \caption{Simulation Parameters for the UAV Situation Case $1$}
\label{Sim_Table_UAV}
 \end{table}

\begin{table}[h]
\begin{tabular}{|p{5cm}|p{8cm}|}
\hline
\vspace{.1mm}USQUE &\vspace{.1mm} $P_a(0)=std^2_{q_0}I_{3\times 3}$, $P_b(0) = std^2_{b_0}I_{3\times 3}$, $P_c(0)=0_{3\times 3}$\\
\hline
\vspace{.1mm}GAME, MEKF, RIEKF &\vspace{.1mm}$P_a(0)=\frac{1}{\vspace{.01mm}std^2_{q_0}}I_{3\times 3}$, $P_b(0) = \frac{1}{\vspace{.01mm}std^2_{b_0}}I_{3\times 3}$, $P_c(0)=0_{3\times 3}$\\
 \hline
 \vspace{.1mm}CGO &\vspace{.1mm}$k_p=1$, $k_I=0.3$\\
 \hline
 \end{tabular}
  \caption{Initial Filter Gain Matrices for the UAV Situation Case $1$}
\label{Sim_Table_tun}
 \end{table}

\begin{figure}[p]
 \includegraphics[scale=.4]{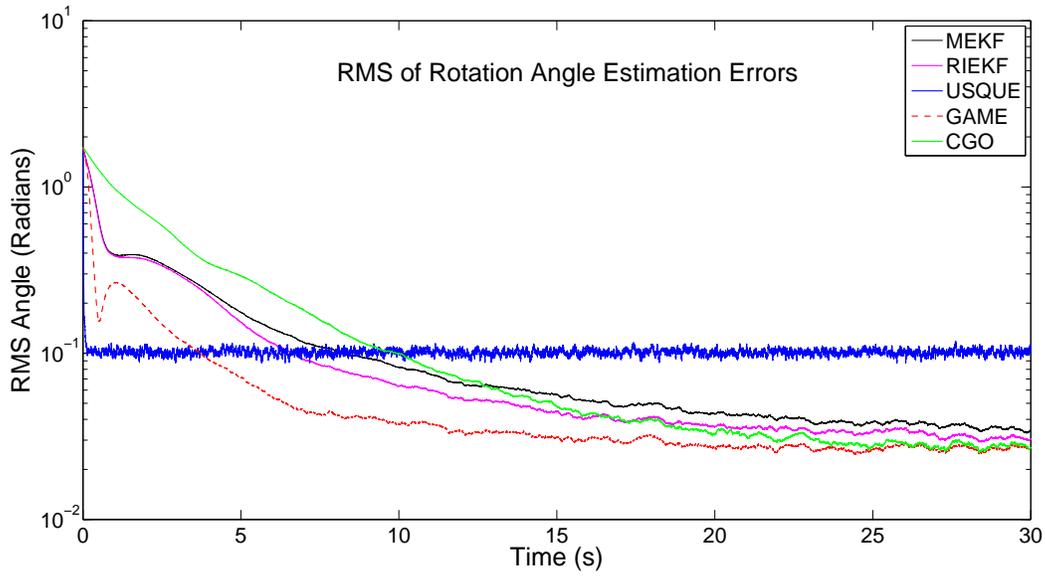}
 \caption{Case 1: The RMS of the estimation error in angle of rotations for a UAV simulation setup. Note that the angle axis is in logarithmic scale.} \label{Sim_UAV_angle}
\end{figure}

\begin{figure}[p]
 \includegraphics[scale=.4]{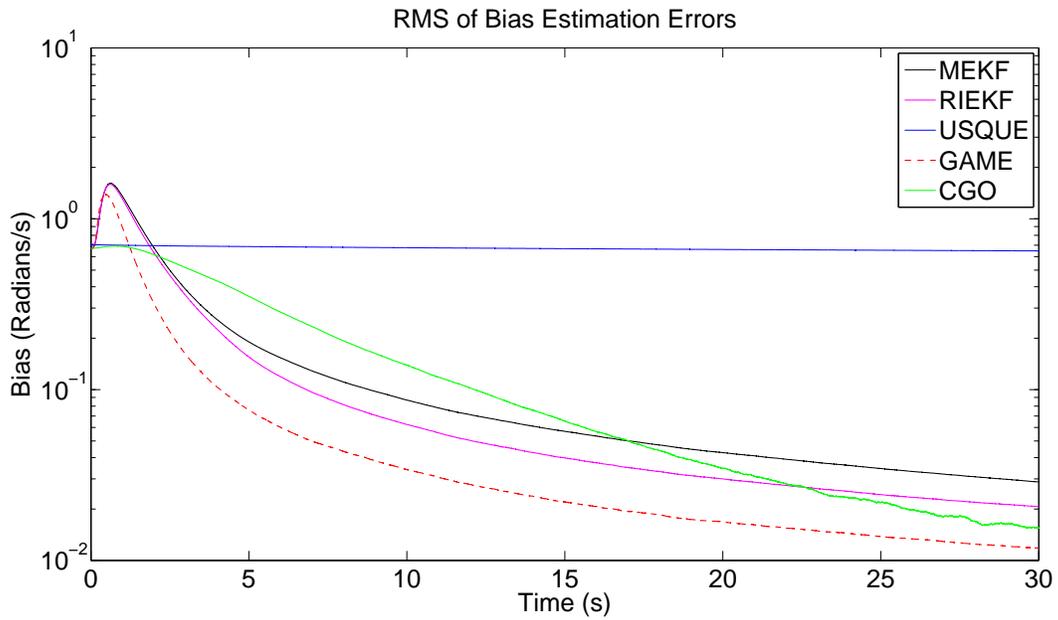}
 \caption{Case 1: The RMS of the bias estimation error for a UAV simulation setup. Note that the bias axis is in logarithmic scale.} \label{Sim_UAV_bias}
\end{figure}

Figures~\ref{Sim_UAV_angle} and~\ref{Sim_UAV_bias} show the performance of the GAME filter compared against the MEKF, the RIEKF, the USQUE and the CGO in the Case 1 experiment.
We have performed a Monte-Carlo simulation and the RMS of the estimation errors of the two filters are shown for $100$ repeats.

\begin{figure}[p]
 \includegraphics[scale=.4]{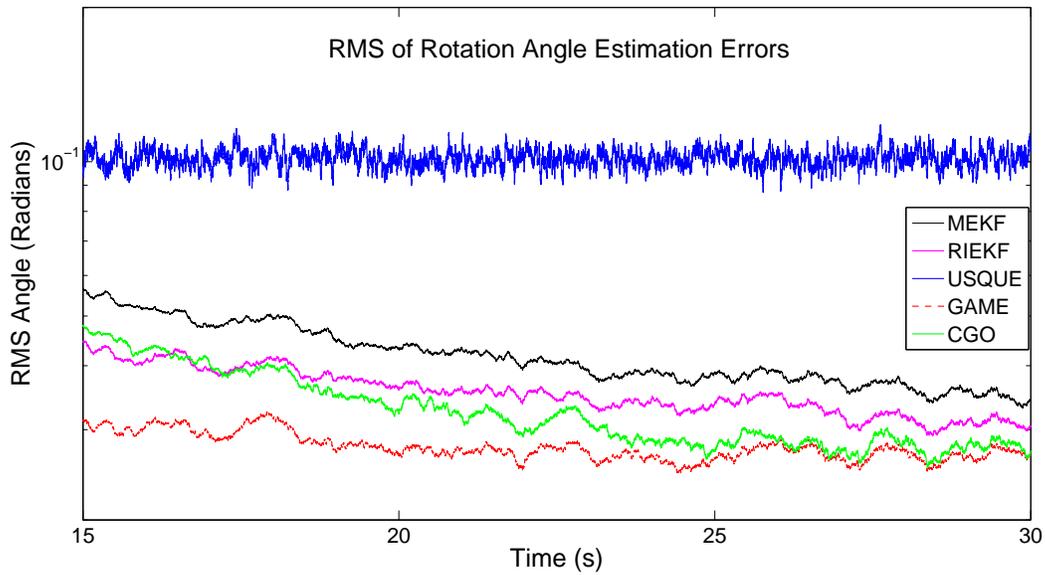}
 \caption{Case 1: The RMS of the estimation error in angle of rotations for a UAV simulation setup. Zoomed on the asymptotic error of the GAME filter, the MEKF and the RIEKF.  Note that the angle axis is in logarithmic scale.} \label{Sim_UAV_angle_Zoom}
\end{figure}

Figures~\ref{Sim_UAV_angle} and the zoomed version (Figure~\ref{Sim_UAV_angle_Zoom}) indicate that the RMS of the rotation angle estimation error of the proposed GAME filter rapidly converges towards zero in the transient period and also maintains the lowest error compared to the rest of the filters in the asymptotic response. 
Figure~\ref{Sim_UAV_bias} shows that the GAME filter also has the lowest asymptotic bias estimation error compared to the bias estimation error of all the other filters.

Note that the initial peak in the angle error of the GAME filter, the MEKF and the RIEKF is associated with the period where the bias estimates of these filters are not accurate enough yet. 
The adaptive nature of these filters is allowing a higher uncertainty in the angle estimates until a reasonable bias estimate is obtained which is then used to achieve an accurate asymptotic angle estimate. 
The bias error of these filters in Figure~\ref{Sim_UAV_bias} is showing the peaking phenomenon that is also seen in a high-gain observer. 
This is not the case for the USQUE which has the fastest angle estimation but the slowest bias estimation. This is due to the fact that the USQUE is setting a high gain for its angle observer and a low gain for its bias observer that leads to the amplified noise on the angle estimation error of the USQUE in Figure~\ref{Sim_UAV_angle}.  
The CGO on the other hand is not an adaptive filter but it has an asymptotically convergent estimation error that is a priori adjusted through the gains $k_p$ and $k_I$.

\begin{figure}[p]
 \includegraphics[scale=.4]{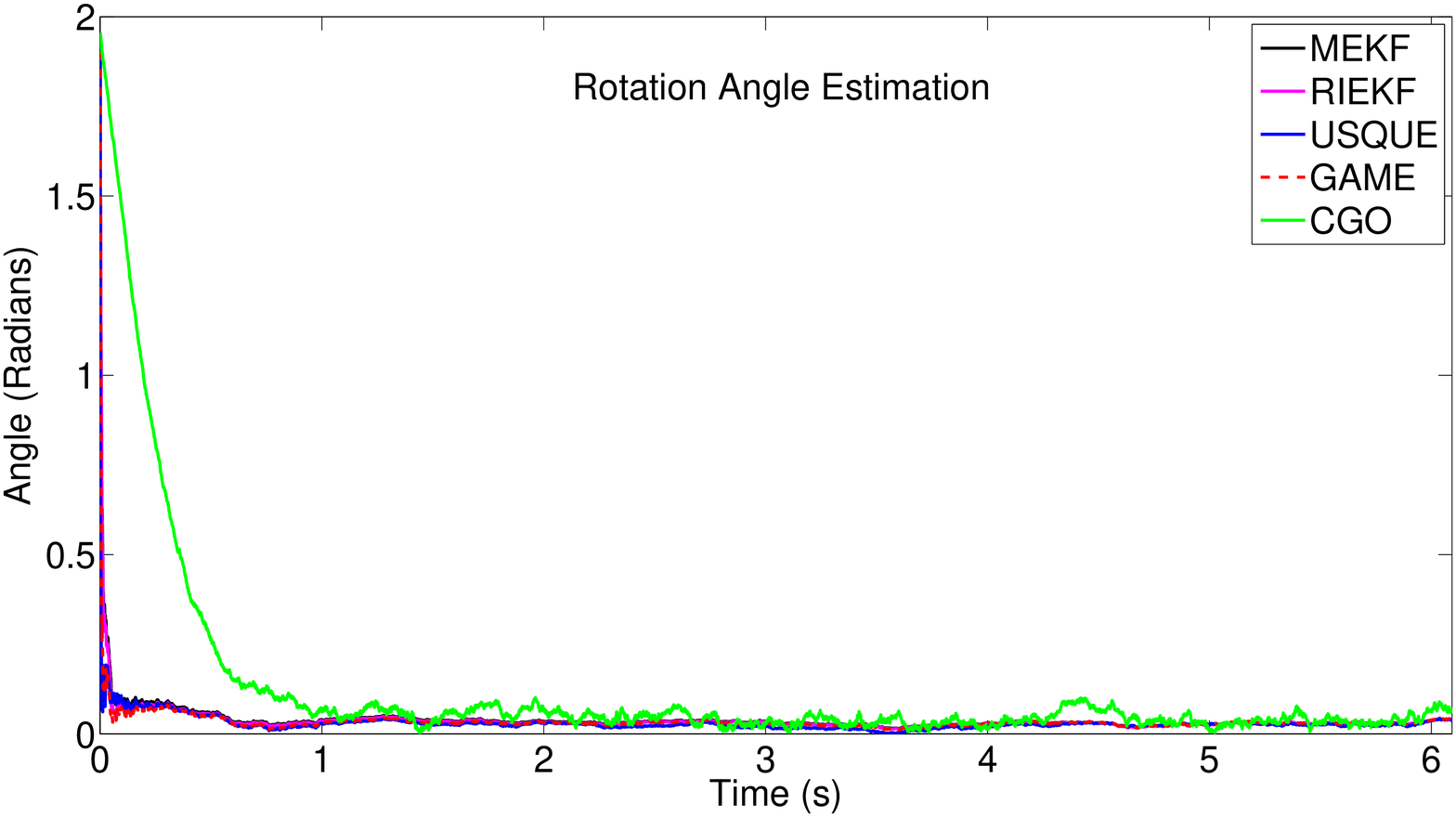}
 \caption{Case 1: The estimation error in angle of rotations for a UAV simulation setup with no bias. Note that in this case it is easily possible to obtain similar convergence behaviour among the filters.} \label{Sim_UAV_nobias}
\end{figure}

 A key aspect of the comparison is the bias estimation part and its effect on the angle estimation performance of the filters. While bias estimation is a practical requirement,  it was observed that excluding the bias and applying proper tuning to the filters leads to little difference in the transient and asymptotic behaviour of the filters. Figure~\ref{Sim_UAV_nobias} demonstrates an instance of the previous simulation where no bias is considered. It was easily possible to find tuning parameters that yield almost identical transient and asymptotic behaviours of all the competing filters except the constant gain observer (CGO).  The CGO in this case has to trade-off between the fast transient behaviour and the asymptotic estimation error. A higher gain results in the faster transient convergence, however, with a larger asymptotic estimation error. This problem is not present in the other filters as their gains are adjusted dynamically by the filter.

Also note that the RIEKF is in fact outperforming the MEKF as was noted in~\cite{GMEKF} too. 
It is interesting that the CGO has the second lowest estimation error with the lowest computational cost. 
Of course the downside of the CGO is that it needs exact tuning depending on the information about the true attitude trajectory that might not always be available a priori. 
The USQUE has the fastest angle convergence to a relatively low error.
However, the noisy asymptotic performance (which might be due to lack of complicated tuning in our experiments), very slow bias estimation and the heavy computational cost of the USQUE compared to the other filters considered makes the USQUE not desirable for the UAV application considered. 
This argument is further investigated in \S~\ref{USQUE_scaling} below.


\subsection{Case 2: Measurement Errors Expected in a Satellite}
In this experiment much smaller measurement error signals are considered as is the case for a satellite attitude filtering problem. 
The simulation parameters are according to the reference~\cite{Crassidis2} and are described in Table~\ref{Sim_Table_Sat}.
Note that the angular velocity input of the attitude kinematics has a much smaller frequency compared to the UAV case as the movement of a satellite is restricted to an earth orbit.

The initial gains of the filters are chosen according to Table~\ref{Sim_Table_Sattun}.
Note that these values are not exactly according to the statistics of the initialization errors of the system, as was the case in our previous experiment. 
This is to avoid singularities that are due to the fact that the numerical values of some simulation parameters are too small and close to the computational limits of the MATLAB programming platform.

\begin{table}[h]
\begin{tabular}{|p{5cm}|p{8cm}|}
\hline
\vspace{.1mm}Time Step &\vspace{.1mm} 0.001 ($s$)\\
\hline
\vspace{.1mm}Simulation Time & \vspace{.1mm}50 ($s$)\\
\hline
\vspace{.1mm}Angle of Rotation Initialization Error  & \vspace{.1mm}$\mathcal{N}\sim(0,60^2)^{\circ}$\\
\hline
\vspace{.1mm}Bias Initialization Error   & \vspace{.1mm}$\mathcal{N}\sim(0,20^2)\frac{^{\circ}}{s}$\\
\hline
\vspace{.1mm}Reference Directions & \vspace{.1mm}$\mathring{y}_1=[1\; 0 \;0]$, $\mathring{y}_2=[0\; 1 \;0]$\\
\hline
\vspace{.1mm}Input signal& \vspace{.1mm}$\Omega = \sin(\frac{2\pi}{150}t)[1\; -1\;\; 1]\frac{^{\circ}}{s}$\\
\hline
\vspace{.1mm}Input error $B_{\Omega}v_{\Omega}$ &\vspace{.1mm} $\mathcal{N}\sim(0,0.31623^2)\frac{\mu\;rad}{s}$\\
\hline
\vspace{.1mm}Bias Variation $B_bv_b$ &\vspace{.1mm} $\mathcal{N}\sim(0,0.031623^2)\frac{n\;rad}{s}$\\
\hline
\vspace{.1mm}Measurement error  $D_iw_i$ &\vspace{.1mm} $\mathcal{N}\sim(0,1^2)^{\circ}$\\
\hline
  \end{tabular}
  \caption{Simulation parameters for a satellite situation}
\label{Sim_Table_Sat}
 \end{table}
\begin{table}[h]
\begin{tabular}{|p{5cm}|p{8cm}|}
\hline
\vspace{.1mm}USQUE &\vspace{.1mm} $P_a(0)=std^2_{q_0}I_{3\times 3}$, $P_b(0) = std^2_{b_0}I_{3\times 3}$, $P_c(0)=0_{3\times 3}$\\
\hline
\vspace{.1mm}GAME, MEKF, RIEKF &\vspace{.1mm}$P_a(0)=\frac{10^{-1}}{\vspace{.01mm}std^2_{q_0}}I_{3\times 3}$, $P_b(0) = \frac{10^{-9}}{\vspace{.01mm}std^2_{b_0}}I_{3\times 3}$, $P_c(0)=0_{3\times 3}$\\
 \hline
  \vspace{.1mm}CGO &\vspace{.1mm}$k_p=10$, $k_I=2$\\
 \hline
 \end{tabular}
  \caption{Initial Filter Gain Matrices for the Satellite Case $2$}
\label{Sim_Table_Sattun}
 \end{table}

\begin{figure}[p]
 \includegraphics[scale=.4]{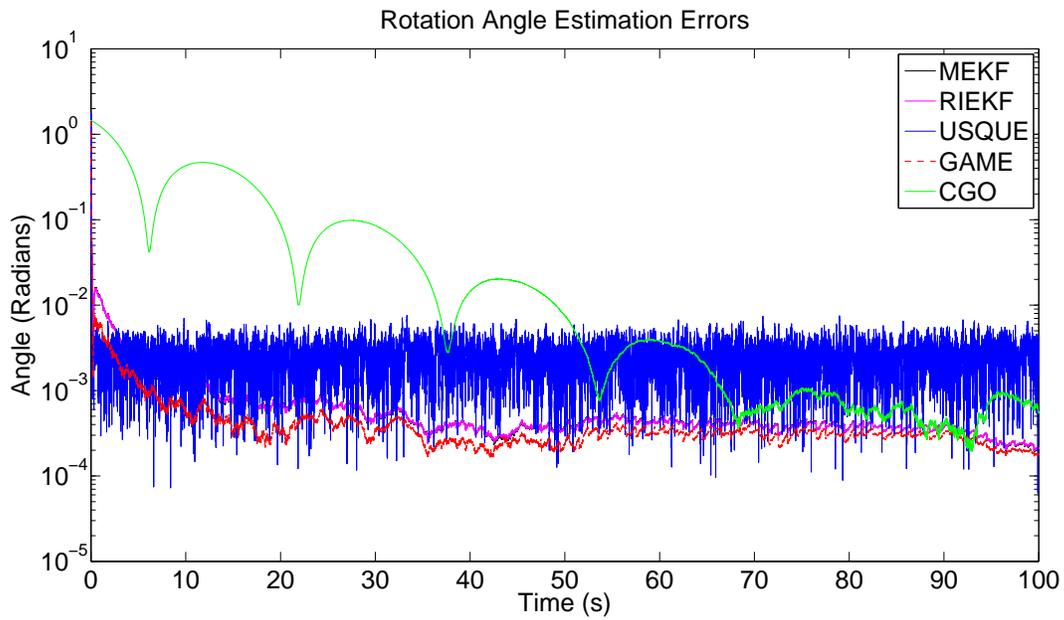}
 \caption{Case 2: The estimation error in angle of rotations for a satellite simulation setup. Note that the angle axis is in logarithmic scale.} \label{Sim_SAT_angle}
\end{figure}

\begin{figure}[p]
 \includegraphics[scale=.4]{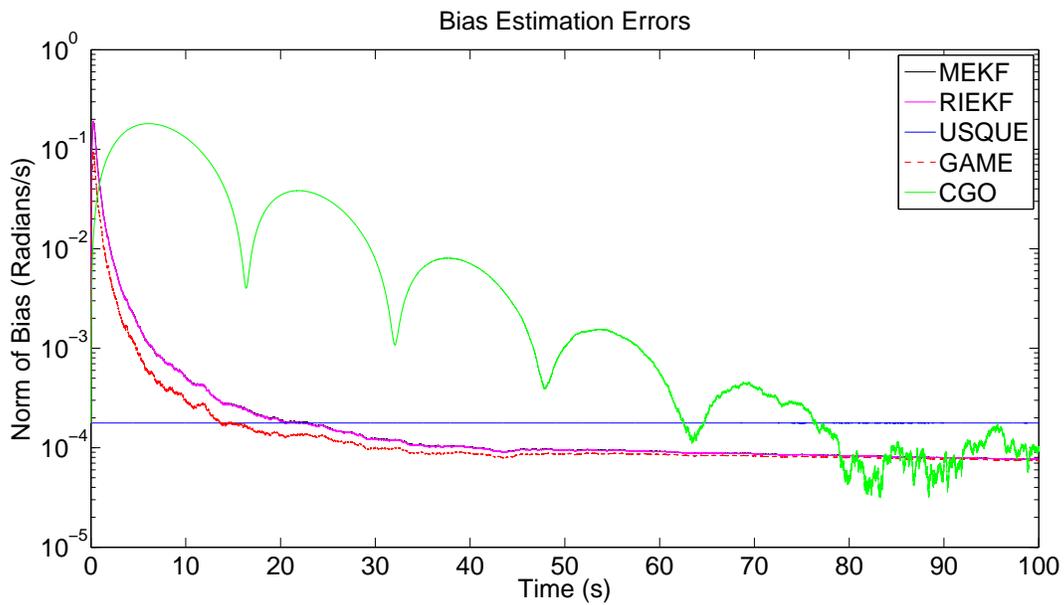}
 \caption{Case 2: The bias estimation error for a satellite simulation setup. Note that the bias axis is in logarithmic scale.} \label{Sim_SAT_bias}
\end{figure}

Figures~\ref{Sim_SAT_angle} and~\ref{Sim_SAT_bias} show the performance of the GAME filter compared against the MEKF, the RIEKF, the USQUE and the CGO in the Case $2$ experiment. 
Note that the figures shown are due to a single repeat of an experiment that is typical for the results seen in more repeats. 
In this case the small frequency of the angular velocity input leads to a slow dynamics of the angle trajectory. 
Due to this slow dynamics and also due to the small measurement errors considered, the estimation errors of all the filters converge towards zero rapidly. 
The GAME filter outperforms the other filters in achieving the lowest asymptotic estimation error. 
The USQUE converges very fast although its asymptotic estimation error is noisy as was the case in the UAV experiment.

\subsection{Gain Scaling}\label{USQUE_scaling}
One can choose different gain scalings for a particular filter. 
A higher scaled gain can result in faster convergence of a filter with the disadvantage of larger asymptotic estimation error.
Depending on the application, one has to trade-off between the transient and the asymptotic performance of a filter. 
The following two examples are demonstrations of this trade-off seen in the results above.

\begin{figure}[p]
 \includegraphics[scale=.4]{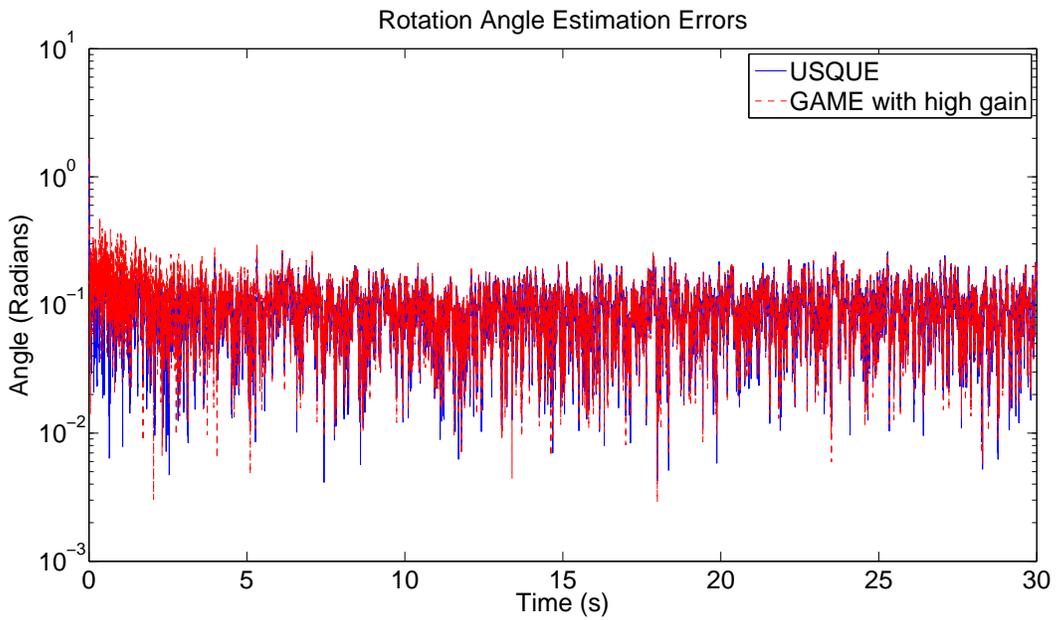}
 \caption{Case 1: The USQUE is compared against the GAME filter when the angle gain of the GAME $P_a$ is multiplied by $30$. Note that the angle axis is in logarithmic scale.} \label{Sim_UAV_USQUE}
\end{figure}

As was apparent in Figures~\ref{Sim_UAV_angle} and~\ref{Sim_UAV_bias}, the USQUE is inherently using a higher angle gain than the other filters. 
In fact, zooming into the asymptotic angle estimation error graph of the USQUE, it is apparent  that the estimation error is approximately $30$ times lager than the estimation error of the other filters (See Figure~\ref{Sim_UAV_USQUE}).
Consider using a scale factor of $30$ multiplying gain $P_a$ of the GAME filter.
As can be seen in Figure~\ref{Sim_UAV_USQUE} the estimation errors of the two filters are now almost identical, confirming that the USQUE algorithm is rendering an undesirable scaling in its angle gain $P_a$ that, although it results in a very fast angle estimation also results in a noisy asymptotic estimation error. 
We have tried to account for this effect in the USQUE algorithm by means of simple tuning. 
However, due to the involved nature of this algorithm (Table~\ref{Sim_Table_USQUE}) it is unclear how to compensate for the clear scaling issue in the gain tuning - this is a clear disadvantage of the USQUE algorithm. 
It is worth noting, that the bias estimation of the GAME filter is still much faster than that of the USQUE indicating the advantage of the GAME filter over the USQUE even in this case.

\begin{figure}[p]
 \includegraphics[scale=.4]{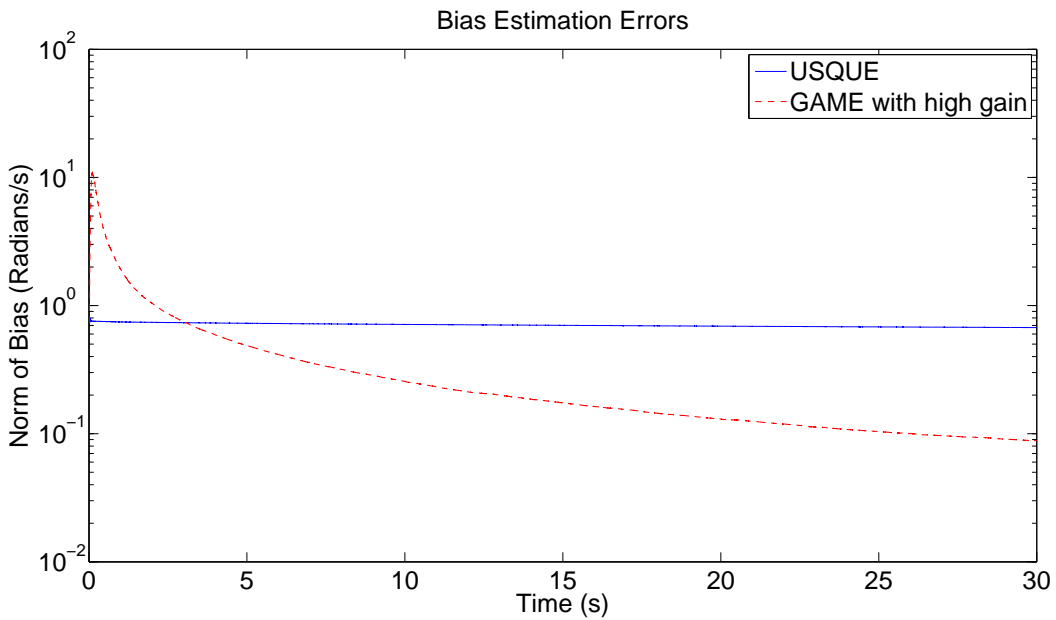}
 \caption{Case 1: The USQUE is compared against the GAME filter when the angle gain of the GAME $P_a$ is multiplied by $30$. Note that the bias axis is in logarithmic scale.} \label{Sim_UAV_USQUE_Bias}
\end{figure}

\begin{figure}[p]
 \includegraphics[scale=.4]{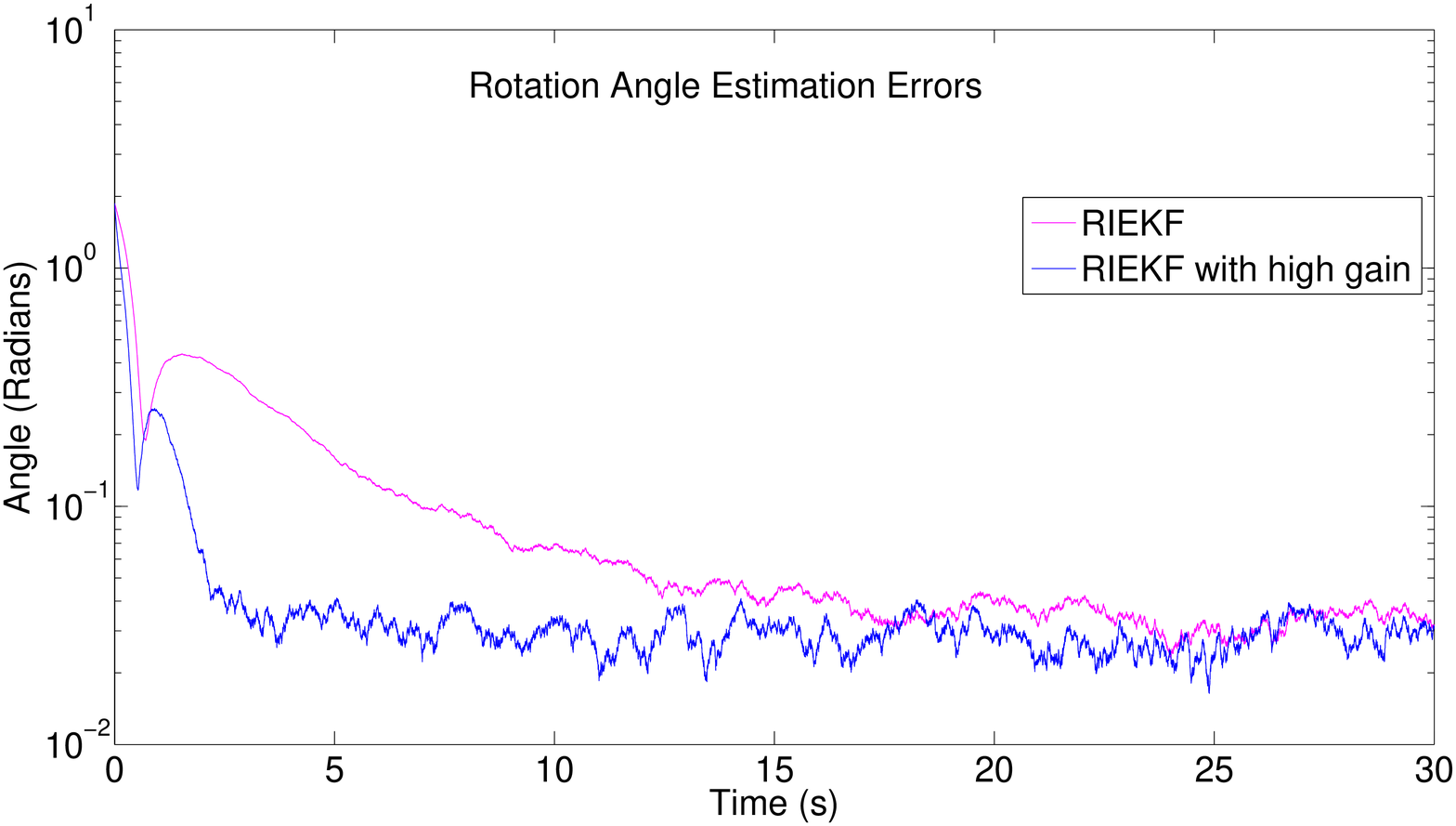}
 \caption{Case 1: The gain difference between the RIEKF in~\cite{GMEKF} and the RIEKF (Table~\ref{Sim_Table_RIEKF}). Note that the angle axis is in logarithmic scale.} \label{Sim_UAV_angle_RIEKF}
\end{figure}

The authors believe that the original published formulation of the RIEKF~\cite{GMEKF} has two typographic errors -- details are provided in the Appendix section.
As a result, the originally published RIEKF has a larger gain than the RIEKF considered here (Table~\ref{Sim_Table_RIEKF}). 
This is investigated in the following simulation comparing the two formulations of the RIEKF.
As can be seen in Figure~\ref{Sim_UAV_angle_RIEKF}, the original RIEKF has a faster decaying convergence error. 
However the asymptotic error of the original RIEKF is approximately two times more noisy than the alternative formulation, confirming the gain difference of the two filters.
This point is important since the relative difference in gain scaling is of the same order as the performance advantage of the RIEKF over the MEKF. 

\clearpage
 \section{Conclusions}\label{Sec_conclusion}

In conclusion, the second-order optimal minimum-energy (or the Geometric Approximate Minimum-Energy, GAME) filter proves to be highly robust both in situations with large measurement errors and fast attitude dynamics, such as the case of a low cost UAV, and also in a situation with small measurement errors and slow attitude dynamics such as in the case of a satellite. In fact in both cases, it was shown in the previous section that the GAME filter outperforms the state-of-the-art attitude filters such as the Multiplicative Extended Kalman Filter MEKF, the Right-Invariant Extended Kalman Filter RIEKF, the UnScented QUaternion Estimator  USQUE and the Constant Gain Observer CGO.

The difference between the MEKF and RIEKF is subtle.  
In low noise situations the two algorithms are essentially equivalent as can be seen from the satellite simulation. 
In the case of high noise it is clear that the more stable noise representation associated with the RIEKF leads to better conditioning of the gain and overall improved performance. 

The USQUE as published is very fast in angle estimation but slow in bias estimation.
Moreover, the asymptotic angle estimation error of the USQUE is very noisy due the high gain of the angle error.
It is clear that the tuning for the USQUE is not optimal, however, due to the complexity of the algorithm it is difficult to see how to improve the gain tuning. 
Given the relative performance of the GAME, MEKF and RIEKF, along with the computational cost of the USQUE, it is unlikely that the effort required to tune the USQUE would be worth pursuing.
A key observation from this study, however, is that comparison of attitude filters must consider both the transient and the asymptotic behaviours and that any filter can produce improved transient behaviour if the filter gain is increased. 

The CGO yields desirable low estimation errors with minimal computational cost. 
However, the gains of the CGO need to be tuned \emph{a priori} and the performance obtained here was heavily based on having used the GAME filter to obtain optimal asymptotic gains for the CGO. 
There is a clear penalty in transient convergence for the CGO compared to the variable gain filters, demonstrating the fundamental advantage of optimal filtering. 
Against this, the constant gain observer does not pose any risk on having an unstable Riccati equation and the computational complexity is very low. 
The CGO clearly remains a very viable filter option in a range of important applications.

\section*{Appendix: RIEKF}\label{Sec_App}\label{app:RIEKF}
 Note that the formulation provided here for the RIEKF is different to the one given in the reference~\cite{GMEKF} in two aspects. 
First, the state error in~\cite{GMEKF} is modeled two times larger, giving the filter formulation a factor of $2$ higher gain. 
Secondly, there seems to be a factor of $\frac{1}{2}$ inconsistency in the $A$ matrix calculation of the RIEKF~\cite{GMEKF} that leads to occasional singularities in the simulation results of the RIEKF. 
The authors believe that both these differences are minor typographical errors in the original paper~\cite{GMEKF}. 
We provide an updated derivation of RIEKF in the following discussion.  

The RIEKF formulation considers the quaternion system model
\begin{equation}\label{App_RIEKF_sys}
 \left\{\begin{array}{l}\dot{q}=\dfrac{1}{2}q\otimes\Omega,\\
         u = \Omega - 2q^{-1}\otimes(B_{\Omega}v_{\Omega})\otimes q +b,\\
         \dot{b}=q^{-1}\otimes(B_Bv_b)\otimes q,\\
         y_i = q^{-1}\otimes(\mathring{y}_i+D_iw_i)\otimes q.\\
         \end{array}\right.
\end{equation}
Note that the state and the output errors are modelled in the inertial frame which is different to the conventional modelling of errors in the body-fixed frame. The RIEKF then is
\begin{equation}\label{App_RIEKF}
 \left\{\begin{array}{l}\dot{\hat{q}}=\dfrac{1}{2}\hat{q}\otimes(u-\hat{b}+2\hat{q}^{-1}\otimes(\sum_iK_q(\mathring{y}_i-\hat{y}_i)\otimes \hat{q}),\\
         \hat{y}_i = \hat{q}\otimes(y_i)\otimes \hat{q}^{-1}, \\
         \dot{\hat{b}}=\hat{q}^{-1}\otimes(\sum_iK_b(\mathring{y}_i-\hat{y}_i)\otimes \hat{q}.\\
         \end{array}\right.
\end{equation}
Consider the errors
\begin{equation}
 \left\{\begin{array}{l}
         \tilde{q} = \hat{q}\otimes q^{-1},\\
         \tilde{b} = q\otimes(\hat{b}-b)\otimes q^{-1}.
        \end{array}
\right.
\end{equation}
The error system is given by
\begin{equation}
 \left\{\begin{array}{l}
         \dot{\tilde{q}} = -\frac{1}{2}\tilde{q}\otimes(\tilde{b})+(\sum_iK_q(\mathring{y}_i-\hat{y}_i)\otimes \hat{q})\otimes\tilde{q}-\tilde{q}\otimes(B_{\Omega}v_{\Omega}),\\
         \dot{\tilde{b}} = +2(B_{\Omega}v_{\Omega})\times \tilde{b}+\tilde{q}^{-1}\otimes(\sum_iK_b(\mathring{y}_i-\hat{y}_i)\otimes \tilde{q}-B_Bv_b+(\tilde{q}^{-1}\otimes(\tilde{u})\otimes \tilde{q})\times \tilde{b},\\
         \tilde{u} = \hat{q}^{-1}\otimes(u-\hat{b})\otimes \hat{q}.
        \end{array}
\right.
\end{equation}
Next, linearize the error system using $\tilde{q}\longrightarrow [1,\frac{1}{2}\delta\tilde{q}]^{\top}$ and $\tilde{b} \longrightarrow \delta\tilde{b}$, and neglect the quadratic terms in noise
and infinitesimal state error similar to~\cite{GMEKF}. Note that the factor of $\frac{1}{2}$ in the linearized quaternion is not considered in the original paper~\cite{GMEKF}. 
\begin{equation}
 \left(\begin{array}{l}
         \dot{\delta\tilde{q}}\\
         \dot{\delta\tilde{b}}\\
         \end{array}\right)=(A-KC)\left(\begin{array}{l}
         \delta\tilde{q}\\
         \delta\tilde{b}\\
         \end{array}\right)-\left(\begin{array}{l}
         B_{\Omega}v_{\Omega}+(\sum_iK_qD_iw_i)\\
         B_bv_b+(\sum_iK_bD_iw_i)\\
         \end{array}\right),
\end{equation}
 where
 \begin{equation}
  A=\left(\begin{array}{cc}
       0 &-I\\
       0 &\tilde{u}_{\times}
      \end{array}\right),\; C = \left( 2(\mathring{y}_i)_{\times}\;\; 0\right), K = -[K_q,K_b]^{\top}.
 \end{equation}

Then similar to the EKF the full filter is realized using
\begin{equation}
 \left\{\begin{array}{l}
         K = PC^{\top}R^{-1},\\
         \dot{P} = AP + PA^{\top} + Q -PC^{\top}R^{-1}CP,\\
         Q = \mbox{diag}(Q_{\omega},\;Q_b).
        \end{array}\right.
\end{equation}
Note that in~\cite{GMEKF} there is a typographic error in the matrix $A$ with an extra factor of $\frac{1}{2}$ multiplying the identity matrix $I$.

If the state noise is considered without a factor of two then
\begin{equation}\label{App_RIEKF_sys_correct}
 \left\{\begin{array}{l}\dot{q}=\dfrac{1}{2}q\otimes\Omega,\\
         u = \Omega - q^{-1}\otimes(B_{\Omega}v_{\Omega})\otimes q +b,\\
         \dot{b}=q^{-1}\otimes(B_Bv_b)\otimes q,\\
         y_i = q^{-1}\otimes(\mathring{y}_i+D_iw_i)\otimes q,\\
         \end{array}\right.
\end{equation}
and the matrix $C$  of the RIEKF is modified to
\begin{equation}
   C = \left( (\mathring{y}_i)_{\times}\;\; 0\right),
 \end{equation}
the version used in Table~\ref{Sim_Table_RIEKF}.
\bibliographystyle{IEEEtran}
\bibliography{ref}

\end{document}